\documentstyle[prl,twocolumn,aps,amsmath,epsfig]{revtex}

\begin{document}
\draft
\twocolumn[\hsize\textwidth\columnwidth\hsize\csname@twocolumnfalse%
\endcsname

\title{Fluctuations of the correlation dimension at metal-insulator
transitions}
\author{E. Cuevas$^1$, M. Ortu\~no$^1$, V. Gasparian$^2$, and
A. P\'erez-Garrido$^3$}
\address{$^1$Departamento de F{\'\i}sica, Universidad de Murcia,
E-30071 Murcia, Spain.\\
$^2$Department of Physics and Geology, California State
University, Bakersfield, CA93311 USA.\\
$^3$Departamento de F{\'\i}sica Aplicada, Universidad Polit\'ecnica 
de Cartagena, E-30202 Murcia, Spain.}

\date{\today}
\maketitle

\begin{abstract}
We investigate numerically the inverse participation ratio, $P_2$, of
the 3D Anderson model and of the power-law random banded matrix (PRBM)
model at criticality. We found that the variance of $\ln P_2$  scales
with system size $L$ as $\sigma^2(L)=\sigma^2(\infty)-A L^{-D_2/2d}$,
being $D_2$ the correlation dimension and $d$ the system dimension.
Therefore the concept of a correlation dimension is well defined in the
two models considered. The 3D Anderson
transition and the PRBM transition for $b=0.3$ (see the text for the
definition of $b$) are fairly similar with respect to all critical 
magnitudes studied.
\end{abstract}

]
\narrowtext

Critical eigenfunctions at metal-insulator transitions show 
multifractality, whose study has been very intensive in the last
two decades \cite{W80,CP86}. Each critical eigenfunction
$\phi_\alpha(\mathbf r)$ can be characterized by a set of inverse
participation ratios (IPR)
\begin{equation}
P^{(\alpha)}_q=\int |\phi_\alpha({\mathbf r})|^{2q}\;d^dr\;,\label{ipr}
\end{equation}
where the index $\alpha$ label the different eigenfunctions and
$d$ is the embedding dimension. In a good metal the IPR scale with size
$L$ as $P^{(\alpha)}_q\propto L^{-d(q-1)}$, while in an insulator
$P^{(\alpha)}_q\propto L^{0}$. Wegner\cite{W80} found, from a
renormalization-group treatment of the metal-insulator transition in
$2+\epsilon$ dimensions, that the average IPR at criticality show an
anomalous scaling of the form
\begin{equation}
P_q\propto L^{-D_q(q-1)}\; ,\label{df}
\end{equation}
being $D_q$ a set of generalized fractal dimensions.

The IPR fluctuations were studied for 2D systems in the framework of
the supersymmetry method\cite{FM95,PA98,M00} and it was found that
the distribution function of $P^{(\alpha)}_q$ normalized to its 
typical value $P_{2}^{\rm (typ)}$ is scale invariant at criticality
and inversely proportional to the squared adimensional conductance
\cite{PA98}. Although the 2D case does not present a true Anderson
transition, the previous result motivated the conjecture that in general
the distribution function of $P^{(\alpha)}_q$ normalized to its typical
value, is universal, i.e., size independent for $L\to\infty$. Accordingly,
it is assumed that the distribution function of $\ln P^{(\alpha)}_q$ is a
universal curve which is just horizontally shifted by changes in $L$.

Recently it was claimed by Parshin and Schober \cite{PS99}, on the 
basis of numerical calculations, that the correlation dimension $D_2$
in the 3D Anderson model at criticality is not well-defined due to
strong  fluctuations in the IPR. For the sizes considered, it seemed
that the standard deviation of the distribution of $\ln P^{(\alpha)}_2$
grows with system size proportionally to $\ln L$. Then, for $L\to\infty$
the correlation dimension, instead of tending to a single value, it would
tend to a universal distribution.

If confirmed, the previous result would impose drastic changes in our
understanding of critical properties at metal-insulator transitions.
Mirlin and Evers \cite{EM00,ME00} addressed this problem and study
theoretically and numerically the fractal properties of the power-law 
random banded matrix (PRBM) model at criticality. They found the
distribution function of the IPR to be scale independent for all the
values of the parameter $b$ characterizing this model (see below).
The model describes a whole family of critical theories parameterized
by $b$, in the same way as the dimensionality labels the different
Anderson transitions. The standard 3D Anderson transition should be
equivalent to a PRBM model with a $b$ of the order of unity, although
this belief is not based on any direct knowledge. They finally claimed
that the disagreement between their results and those of Ref.
\cite{PS99} is due to the small system sizes used in  \cite{PS99}
and not to the different models employed.

Our aim is to perform a careful statistical analysis on the data from
numerical calculations of both the 3D Anderson model and the PRBM 
model at criticality using system sizes larger than in previous
calculations \cite{PS99,EM00}. We want to study the system size
dependence of the fluctuations of the IPR and to elucidate: 
i) whether the correlation dimension at the Anderson transition is 
well-defined or alternatively presents an scale invariant distribution, 
ii) whether this transition is equivalent or not to a PRBM model
at criticality. 

We first consider the standard Anderson model on a 3D simple cubic
lattice, represented by a tight-binding Hamiltonian with matrix 
elements of the form
\begin{equation}
H_{ij} = \epsilon_i \delta_{ij}+t_{ij}
\label{hamil}\;,
\end{equation}
where $i,j$ denote lattice sites. The diagonal (site) energies are
randomly distributed with constant probability in the interval
$-W/2<\epsilon_i <W/2$, and the off-diagonal elements $t_{ij}$ are 
taken equal to unity for nearest neighbors, which sets the energy 
scale, and to zero otherwise. 

The PRBM model, introduced in Ref.\ \cite{Mirlin}, describes a 1D sample
with random long-range hopping. It can approximately represent a variety
of physical systems from an integrable billiard with a Coulomb scattering
center \cite{AL97}, to the Luttinger liquid at finite temperatures
\cite{K99,KS00}. The model is represented by real symmetric matrices whose
entries are randomly drawn from a normal distribution with zero mean and a
variance depending on the distance of the matrix element from the diagonal
\begin{equation}
\left\langle (H_{ij})^2\right\rangle =\frac{1}{1+(|i-j|/b)^{2\alpha}} 
\label{hamil2}\;.
\end{equation}
This model was shown to undergo a sharp transition at $\alpha=1$ from
localized states for $\alpha>1$ to delocalized states for $\alpha<1$.
This transition is supposed to be similar to an Anderson metal-insulator
transition, presenting multifractality of eigenfunctions and non-trivial
spectral compressibility at criticality. The parameter $b$ determines the
critical dimensionless conductance and so establishes the character of the
transition. For $b=1$ the nearest level spacing distribution differs 
from the typical one at the 3D Anderson metal-insulator transition 
\cite{VB00}. Recent calculations by us have obtained different diverging
exponent for the correlation length as the transition is approached from
below and from above \cite{CG01}.

We obtain the eigenfunctions and eigenvalues of the Hamiltonian 
matrix by numerical diagonalization. In the case of the Anderson model,
we use techniques for large sparse matrices, in particular a Lanczos
tridiagonalization without reorthogonalization method \cite{CW85}, while
for the PRBM case we employ standard diagonalization subroutines, since
we have to deal with full matrices.
For the Anderson model, the system size varies between 5 and 40, and we
consider a small energy window $(-1,1)$ around the center of the band.
We take for the critical disorder the value $W_{\rm c}=16.5$. In the PRBM
case, the system size ranges between $L=100$ and 15000 and the 
energy window considered is  $(-0.4,0.4)$. Reducing the width of the
previous windows do not alter the results. 
The number of random realizations is such that the number of
states included for each $L$ is roughly equal to $3\times 10^5$, except
for $L=40$ in the Anderson model where this number is $2\times 10^4$.
In order to reduce edge effects, we use periodic boundary conditions in
all cases considered.

To elucidate whether the correlation dimension $D_2$ possesses a well
defined single value or alternatively corresponds to a distribution,
we have calculated the IPR for the wavefunctions in the energy window
considered for many disorder realizations. For each $L$ we obtain the
distribution function of $\ln P^{(\alpha)}_2$, since this is a well
behaved self-averaging magnitude. At the end of the paper, we will also
discuss the distribution of $P^{(\alpha)}_2$. In Fig.\ 1 we show the
evolution of the distribution function $F(\ln P^{(\alpha)}_2)$ with $L$
for the 3D Anderson model. The system sizes drawn are $L=5$, 6, 8, 10,
12, 15, 17, 20, 25, 30 and 40, from right to left. It is clear that
$F(\ln P^{(\alpha)}_2)$ changes with size. As $L$ increases  this
distribution becomes wider and the height of its peak smaller. This is
in qualitative, but
\begin{figure}
\epsfxsize=\hsize
\begin{center}
\leavevmode
\epsfbox{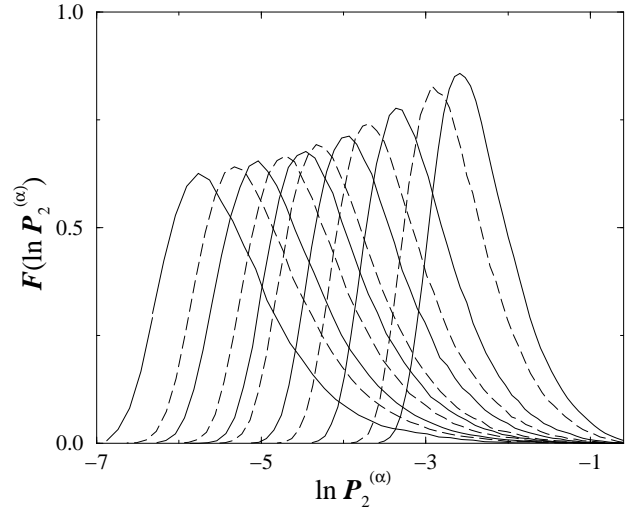}
\end{center}
\caption{Distribution function $F(\ln P^{(\alpha)}_2)$ for the 3D
Anderson model on a logarithmic scale for $L=5$, 6, 8, 10, 12, 15,
17, 20, 25, 30 and 40, from right to left.}
\label{fig1}
\end{figure}
not quantitative, agreement with the results of
Ref.\ \cite{PS99}.

We characterize the previous distribution by its average value, 
$\langle\ln P^{(\alpha)}_2\rangle$, and its variance, 
$\sigma^2(L)={\rm var}(\ln P^{(\alpha)}_2)$.
This variance increases with $L$ and seems
to saturate at a constant value, as one can implicitly appreciate
from the peak of the distributions. 
We try a fit of the form 
\begin{equation}
\sigma^2(L)=\sigma^2(\infty)-A L^{-\gamma}\;,\label{e1}
\end{equation}
with $\sigma^2(\infty)$, $A$ and  $\gamma$ being three adjustable
parameters. We found that the exponent $\gamma$ was always very close
to the correlation dimension divided by $2d$. Thus, we fix
\begin{equation}
\gamma=\frac{D_2}{2d}\;,\label{irre}
\end{equation}
and keep only two free parameters. In the language of scaling theory,
this exponent characterizes the behavior of the {\it irrelevant} length.
Our assumption that the irrelevant exponent $\gamma$ is equal to
$D_2/2d$ properly interpolates between two known limiting cases of the
PRBM model.
For $b\ll 1$, one can get from Eqs.\ (\ref{e1}) and (\ref{irre})
the $b\ln L$ correction predicted in \cite{ME00}.
Similarly, for $b\gg 1$, the exponent $\gamma$ in Eq.\ (\ref{irre}) 
tends to $1/2$, which implicitly coincides with the results of 
Ref.\ \cite{ME00} for this regime.

In Fig.\ 2 we represent
on a log-log scale $\sigma^2(\infty)-\sigma^2(L)$ as a function of $L$ for
the 3D Anderson transition. The fitted values of the free parameters are 
$\sigma^2(\infty)=1.09$ and $A=1.24$. The slope of the straight line 
has not been fitted and corresponds to $D_2/2d$, where $D_2=1.4$
has been obtained from the inset of Fig.\ 2. In this inset we plot
$\langle\ln P^{(\alpha)}_2\rangle$ versus $\ln L$. The straight line is
a linear fit to the data and its slope is equal to $D_2$.
The value $\sigma^2(\infty)=1.09$ found is in good agreement with the
conjecture $\sigma^2(\infty)\approx 1$ \cite{FM95}.

\begin{figure}
\begin{center}
\leavevmode
\epsfbox{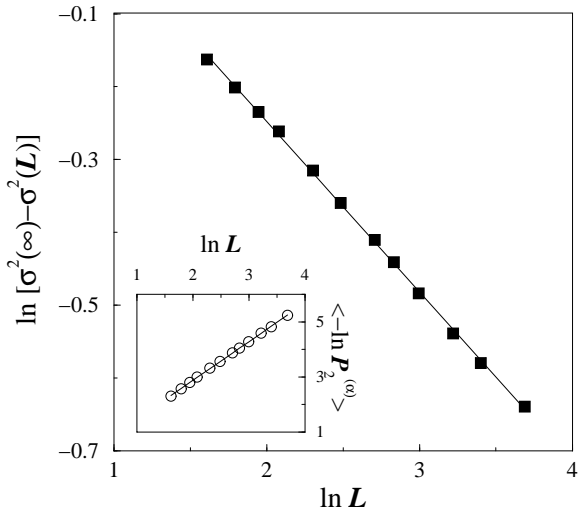}
\end{center}
\caption{$\sigma^2(\infty)-\sigma^2(L)$ as a function of $L$ for the
Anderson transition on a log-log scale. $L$ ranges between 5 and 40.
The straight line is a linear fit to Eq.\ (\ref{e1}), and the slope
is obtained from the inset. Inset: $\langle\ln P^{(\alpha)}_2\rangle$
versus $\ln L$. The straight line is a linear fit to the data and its
slope is equal to the (average) correlation dimension $D_2$.}
\label{fig2}
\end{figure}

These results are in contradiction with those of  Parshin and 
Schober \cite{PS99}, who found a linear increase of $\sigma(L)$ with
$\ln L$. We have checked that $\sigma(L)$ versus $\ln L$ do not follow
a linear behavior.
We believe that the disagreement is due to two reasons. Firstly, the
relatively small system sizes employed in Ref. \cite{PS99}, and secondly 
the use of different estimates of an effective IPR, instead of directly 
studying the width of the $\ln P^{(\alpha)}_2$ distribution. 
On the other hand, Mirlin and Evers
\cite{ME00} already obtained for the PRBM model similar results to ours. 

Eq.\ (\ref{irre}) may be also valid for other transitions, like the
integer quantum Hall transition, where the irrelevant exponent $y$
--which is the same as our $\gamma$-- is known to be $y=0.4\pm 0.1$
and $D_2=1.48$. These values are consistent with Eq.\ (\ref{irre}).
Polyakov \cite{Po99} and Evers and Brenig \cite{EB98} already obtained
a relation between the irrelevant exponent $y$ and the correlation
dimension $D_2$, but our result seems to fit more naturally the
reported values of $y$.

Although $D_2$ is well defined in the macroscopic limit, it is clear from
Fig.\ 2 that the results of numerical calculations with present computers
will drastically depend on the particular definition of correlation
dimension adopted. For the sizes available the distributions of
$\ln P^{(\alpha)}_2$ are so wide that one can obtain very different
numerical values of $D_2$ depending on the method of calculation, as
pointed out in Ref.\ \cite{PS99}, but this does not imply at all the
existence of a distribution of correlation dimensions in the macroscopic
limit.

We have now performed similar calculations for the PRBM model, considering
several values of the parameter $b$ in the range where the transition
could be similar

\begin{figure}
\begin{center}
\leavevmode
\epsfbox{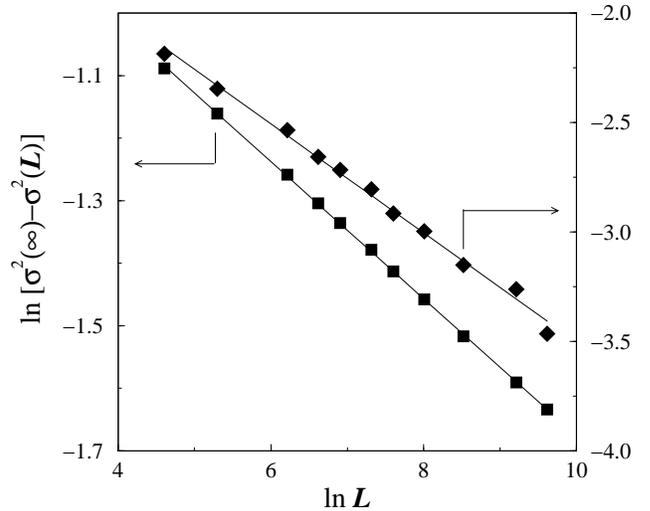}
\end{center}
\caption{$\sigma^2(\infty)-\sigma^2(L)$ as a function of $L$ on a 
log-log scale for the PRBM transition with $b=0.1$ (squares) and 
0.3 (diamonds). Solid lines are fits to Eq.\ (\ref{e1}).}
\label{fig3}
\end{figure}

\noindent to the 3D Anderson transition. As regards the IPR
fluctuations, the results for the PRBM model follow the same behavior,
given by Eq.\ (\ref{e1}), as the Anderson transition.
In order to show this, we plot in Fig.\ 3 the same quantity as in
Fig.\ 2, but for the PBRM model with $b=0.1$ (squares) and with
$b=0.3$ (diamonds). The data in all cases are well fitted by
straight lines, whose slopes are equal to $D_2/2d$. 
The fitted parameters for the cases shown in Fig.\ 3 are  
$\sigma^2(\infty)=0.55$ and $A=0.56$ for $b=0.1$, and 
$\sigma^2(\infty)=0.33$ and $A=0.35$ for $b=0.3$.
The values of $D_2$ have been obtained by the same procedure as for the
Anderson model and are equal to 0.21 and 0.48 for $b=0.1$ and 0.3,
respectively. For $b=1$, $\sigma^2(L)$ is practically constant, 
explaining the scale-invariance claim  made in 
Refs.\ \cite{EM00,ME00}. For $b>1$, Eq.\ (\ref{e1}) is still valid, but
$A$ changes sign. In other words, now the system approaches the asymptotic
value $\sigma^2(\infty)$ from above.

Now we focus on the behavior of the asymptotic value $\sigma^2(\infty)$
as a function of $b$, which is summarized in Fig.\ 4. We were not able
to fit all the data (including large values of $b$ not shown in the
figure) with a simple function. Power laws of $1/b$ with different 
exponents are able to fit extended parts of the curve. For example,
the dashed line corresponds to a $1/b$ dependence and fits fairly
well the intermediate regime, $b\sim 1$, in which we are most interested. 
For $4<b<12$, the results are in reasonable agreement with the $1/b^2$
prediction of Evers and Mirlin \cite{EM00}, based on the fluctuations of
the 2D case and taking into account that the dimensionless conductance is
proportional to $b$.

The distribution of $z\equiv P^{(\alpha)}_2/ P_{2}^{\rm (typ)}$, where the
typical IPR $P_{2}^{\rm (typ)}$ is chosen as the median of the distribution
\cite{Sh86}, is very different from the distribution of
$\ln P^{(\alpha)}_2$ due

\begin{figure}
\begin{center}
\leavevmode
\epsfbox{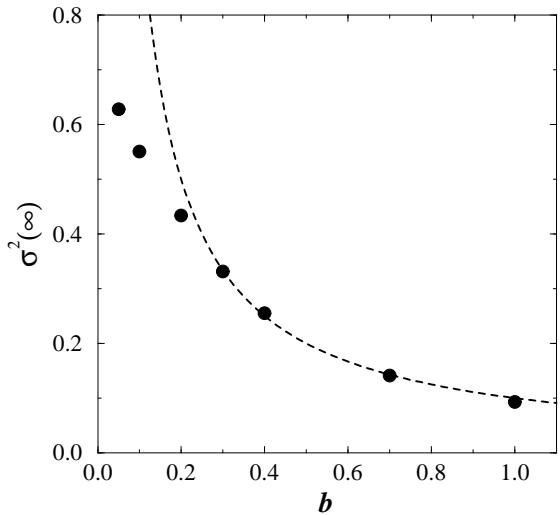}
\end{center}
\caption{Asymptotic value  of $\sigma^2(L)$ as a function of $b$. The
dashed line is proportional to $1/b$.}
\label{fig4}
\end{figure}

\noindent to the their strong asymmetries and long tails.
We have studied the size dependence of the variance $\sigma_z^2$ of $z$. 
For $b\ll 1$ we have analytically calculated this dependence using
the renormalization group method of Levitov \cite{Le99}. One combines two
blocks of size $L$ to form a block of size $2L$. Some states are assumed
to remain practically unchanged in this process, while others combine
to form a more extended state. From the evolution equation describing
the behavior of the distribution function, one can obtain $D_2$
in terms of the parameters of the model \cite{ME00}. We extended the
procedure to obtain the following equation for the variance
\begin{equation}
\sigma_z^2(2L)=2^{-D_2/8}\,\sigma_z^2(L)+\tfrac{b}{2}\,2^{-2D_2}\;.
\label{e2}
\end{equation}
We have checked that for $b<0.3$ Eq.\ (\ref{e2}) fits the variance
of $z$ fairly well. On the other hand, the numerical calculations for
larger values of $b$ and for the 3D Anderson transition do not follow
this behavior. Eq. (\ref{e2}) predicts a power law approach of
$\sigma_z^2(L)$ to its asymptotic value with an exponent $-D_2/8$, in
contrast with the exponent $-D_2/2d$ observed for the variance of
$\ln P^{(\alpha)}_2$.

We have compared the Anderson model with the PRBM model for several
values of $b$ through different critical parameters. The case $b=0.3$
is very similar to the Anderson transition, presenting practically the
same values for all the critical magnitudes studied.
The correlation dimension divided by the embedding dimension is equal
to $1.4/3=0.47$ for the Anderson model and to 0.49 for  the PRBM model. 
Correspondingly, the scaling with system size of the fluctuations of
$\ln P^{(\alpha)}_2$ is rather similar. The asymptotic value of
these fluctuations for the PRBM model is the same as for the Anderson
model divided by $d$. Also, the normalized variance of the nearest level
spacing at criticality \cite{C99,Me91}, which characterizes
the intermediate statistics at the transition \cite{SS93}, is 0.18 for
the Anderson model and 0.19 for the PRBM model with $b=0.3$.

Following the comparison between the two models, we have finally 
analyzed the large values tail of the distribution of $z$. Mirlin and
Evers \cite{EM00,ME00} predicted for the PRBM model a power-law tail
$F(z)\propto z^{-1-x_2}$, where $x_2$ depends on the transition considered. 
For example,  $x_2$ is equal to 4.2 for $b=1$ and 2.1 for  $b=0.3$.
We have checked that this prediction is also obeyed by the Anderson
model, for which we obtained a value of  $x_2$ equal to 1.6, in relative 
agreement with the PRBM model for $b=0.3$.

In summary, we found that the variance of the fluctuations of 
$\ln P^{(\alpha)}_2$ tends to an asymptotic value, implying a 
well-defined correlation dimension. For the intermediate
regime of the PRBM model, $\sigma^2(\infty)$ is  proportional to $1/b$. 
For the Anderson transition and the PRBM model at criticality, the 
variance tends to its asymptotic value as a power law $L^{-D_2/2d}$.
Thus, the irrelevant length scale diverges with an exponent $D_2/2d$,
a result that may be extended to other transitions like the integer
quantum Hall transition.
The behavior of the distributions of $\ln P^{(\alpha)}_2$ and of 
$P^{(\alpha)}_2/P_{2}^{\rm (typ)}$ is different.
We checked that the 3D Anderson transition and the PRBM transition for
$b=0.3$ are similar for all the critical magnitudes analyzed: the
correlation dimension, the scaling of the fluctuations of 
$\ln P^{(\alpha)}_2$, the exponent characterizing the long $z$ tail 
of the distribution $F(z)$ and the normalized nearest level variance.

We would like to thank I.V. Lerner and A. Mirlin for useful discussions,
and the Spanish DGESIC, project numbers 1FD97-1358 and BFM2000-1059, for
financial support.


\end{document}